\documentclass[10pt,a4paper,notitlepage]{article}
\usepackage[utf8]{inputenc}
\usepackage[T1]{fontenc}
\usepackage{multicol} % columns
\usepackage[margin=0.75in]{geometry} % page margin
\usepackage[english]{babel}
\usepackage{csquotes}
\usepackage{adjustbox}
\usepackage{titlesec} % title
\usepackage{float} % needed for figures in two columns
\usepackage{makecell} % cell in tabular
\usepackage{tabularx}
\usepackage{amssymb} % real and natural numbers
\usepackage{siunitx} % math symbol
\usepackage{amsmath} % math
\titleformat*{\section}{\large\bfseries} % section
\titleformat*{\subsection}{\normalsize\bfseries} % subsection
\titleformat*{\subsubsection}{\normalsize\bfseries} % subsubsection
\usepackage[backend=biber, sorting=none, style=nature]{biblatex} % references
\renewcommand{\cite}[1]{\supercite{#1}}
\usepackage[usenames,dvipsnames,svgnames,table]{xcolor}
\usepackage{tcolorbox}
\usepackage{enumitem}
\usepackage{booktabs}
\usepackage{setspace}
\usepackage[shortcuts]{extdash}
\DeclareSIUnit\angstrom{\text{Å}}

\usepackage{hyperref}
\hypersetup{
    colorlinks,
    linkcolor={black},
    citecolor={black},
    urlcolor={black}
}

\begin{document}
\begin{center}

\LARGE \textbf{Structure-based drug design with geometric deep learning} \bigskip

\bigskip
\normalsize {Clemens Isert}$^{1, \dagger}$, {Kenneth Atz}$^{1,\dagger}$ \& {Gisbert Schneider}$^{1,2,*}$ \bigskip

\small $^{1}$ETH Zurich, Department of Chemistry and Applied Biosciences, Vladimir-Prelog-Weg 4, 8093 Zurich, Switzerland. \par
\small $^{2}$ETH Singapore SEC Ltd, 1 CREATE Way, $\#$06-01 CREATE Tower, Singapore, Singapore. \par
$\dagger$ \small These authors contributed equally to this work. \par
\small $*$ To whom correspondence should be addressed. \par
\small E-mail: gisbert@ethz.ch \par
\end{center}

\begin{spacing}{1.0}

\begingroup 
\begin{abstract}
\noindent
Structure-based drug design uses three-dimensional geometric information of macromolecules, such as proteins or nucleic acids, to identify suitable ligands. Geometric deep learning, an emerging concept of neural-network-based machine learning, has been applied to macromolecular structures. This review provides an overview of the recent applications of geometric deep learning in bioorganic and medicinal chemistry, highlighting its potential for structure-based drug discovery and design. Emphasis is placed on molecular property prediction, ligand binding site and pose prediction, and structure-based \textit{de novo} molecular design. The current challenges and opportunities are highlighted, and a forecast of the future of geometric deep learning for drug discovery is presented.
\end{abstract}
\endgroup

\begin{multicols}{2}
\raggedcolumns

%%%%%%%%%
%%%%%%%%%
%%%%%%%%%
%%%%%%%%%

\section{Introduction}

% structure-based drug design and geometric deep learning
Structure\-/based drug design is based on methods that leverage three\-/dimensional (3D) structures of macromolecular targets, such as proteins and nucleic acids, for decision-making in medicinal chemistry.~\cite{gubernator1998structure,anderson2003process} Structure-based modeling is well established throughout the drug discovery process, aiming to rationalize non\-/covalent interactions between ligands and their target macromolecule(s).~\cite{bissantz2010medicinal} The questions addressed with structure-based approaches include molecular property prediction, ligand binding site recognition, binding pose estimation, as well as \textit{de novo} design.~\cite{bleicher2003hit, sledz2018protein, atz2023machine, sadybekov2022synthon} For such tasks, detailed knowledge of the 3D structure of the investigated macromolecular surfaces and ligand\-/receptor interfaces is essential. Recently, an emerging concept of neural\-/network\-/based "artificial intelligence", geometric deep learning, has been introduced to solve numerous problems in the molecular sciences, including structure\-/based drug discovery and design.~\cite{atz2021geometric} \\

Geometric deep learning is based on a neural network architecture that can incorporate and process symmetry information.~\cite{bronstein2021geometric} Effective learning on three\-/dimensional (3D) graphs has become one of the primary functions of molecular geometric deep learning.~\cite{unke2021spookynet, unke2021se, satorras2021n, christensen2021orbnet, nippa2022enabling} Such methods, which were initially limited to small molecules, are now increasingly applied to macromolecules for structure-based drug design.~\cite{ganea2021independent,unke2022accurate} \\

%% Geometry aware neural networks are applications built on this theory. 
%% Language in clm figure

% content
This review provides a concise overview of geometric deep-learning methods for structure-based drug design and seeks to forecasts future developments in the field. We focus on methods that use 3D macromolecular structure representations developed for rational drug design, emphasizing the most recent developments in both predictive and generative deep-learning methods for structure-based molecular modeling. The most relevant representations of 3D protein structures and the essential symmetry operations for geometric learning are discussed (Figure~\ref{fig:representation_overview}). Then, the most recent developments in the field are addressed (Figure~\ref{fig:classification_overview}), namely, (i) molecular property prediction (\textit{e.g.}, binding affinity, protein function, and pose scoring, Section~\ref{sec:property_prediction}), (ii) binding site and interface prediction (\textit{e.g.}, small molecule binding sites and protein\-/protein interfaces, , Section~\ref{sec:binding_site}), (iii) binding pose generation and molecular docking (\textit{e.g.}, ligand\-/protein and protein-protein binding, Section~\ref{sec:docking}), and (iv) the structure\-/based \textit{de novo} design of small-molecule ligands (Section~\ref{sec:denovo}).  

\subsection{Molecular representation}
\label{sec:molecular_representations}

\begin{figure*}[th!]
\centering
\includegraphics[width=\linewidth]{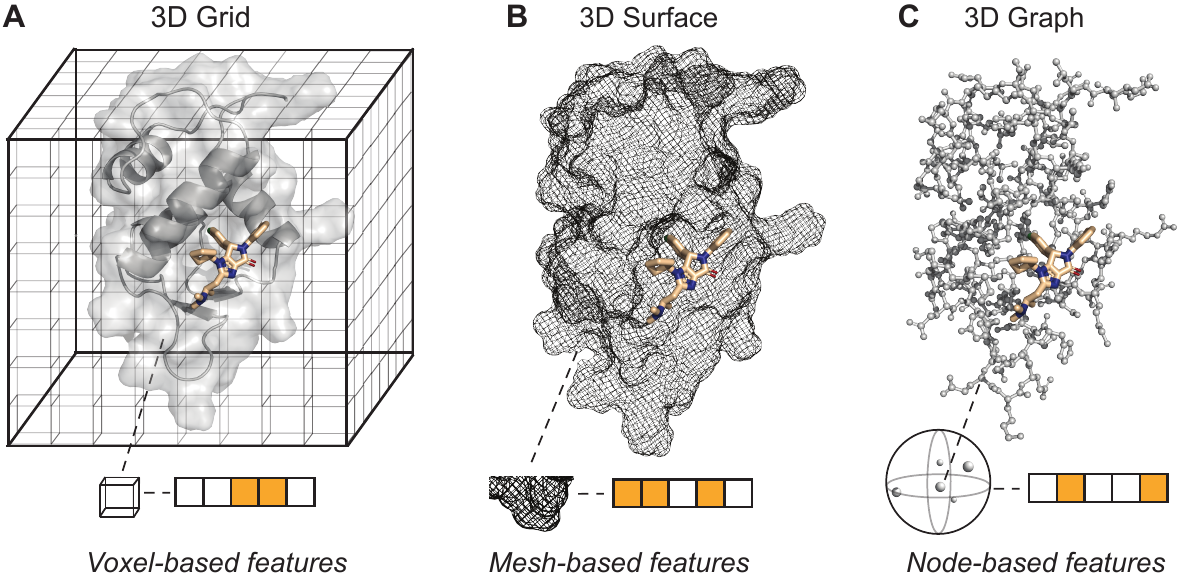}
\caption{Overview of 3D representations of macromolecular structures used for geometric deep learning, illustrated for the human MDM2 protein (PDB-ID 4JRG ~\cite{ding2013discovery}). Ligand DF-1 was docked into the active site of MDM2. The highest-ranking docking pose is shown. DF-1 was automatically generated using a novel geometric deep-learning method (see Section \ref{section:df} for details). The white and orange squares illustrate the feature vectors corresponding to the different representations. (\textbf{A}) 3D grid. (\textbf{B}) 3D surface. (\textbf{C}) 3D graph. For visual clarity, only covalent edges are shown.}
\label{fig:representation_overview}
\end{figure*}

The representation of the macromolecular structure depends on the machine learning task and the chosen architecture. The three most prevalent macromolecular representations described in the recent literature are grids, surfaces, and graphs (Figure~\ref{fig:representation_overview}). These three representations have unique geometries and symmetries:~\cite{bronstein2021geometric} 
\begin{itemize}
    \item \textit{3D grids} are defined by a Euclidean data structure consisting of voxels in 3D space. This Euclidean geometry features the individual voxels of the grid with a fixed neighborhood geometry, \textit{i.e.} (i) each voxel has an identical neighborhood structure (defined by the number of neighbours and distances) and is indistinguishable from all other voxels from a purely structural perspective, and (ii) the voxels have a fixed order that is defined by the spatial dimensions of the grid.
    \item \textit{3D surfaces} consist of polygons (faces) that describe the 3D arrangement of the mesh coordinates ("mesh space"). The polygons can be distinguished according to their chemical features and geometric features defined by the local geometry of the mesh. 
    \item \textit{3D graphs} are defined by a non-Euclidean data structure consisting of nodes (represented by the individual atoms) and their edges, which are defined by the neighboring nodes, \textit{e.g.}, through a certain distance cutoff or $k$-nearest neighbor assignment. The non-Euclidean geometry of graphs originates from a non-consistent neighborhood structure of the individual nodes, \textit{i.e.} each node can have a different number of neighbors with edges defined by different spatial distances. There is no general ordering of nodes and edges.  
\end{itemize}

\subsection{Symmetry}
\begin{figure*}[ht!]
\centering
\includegraphics[width=\linewidth]{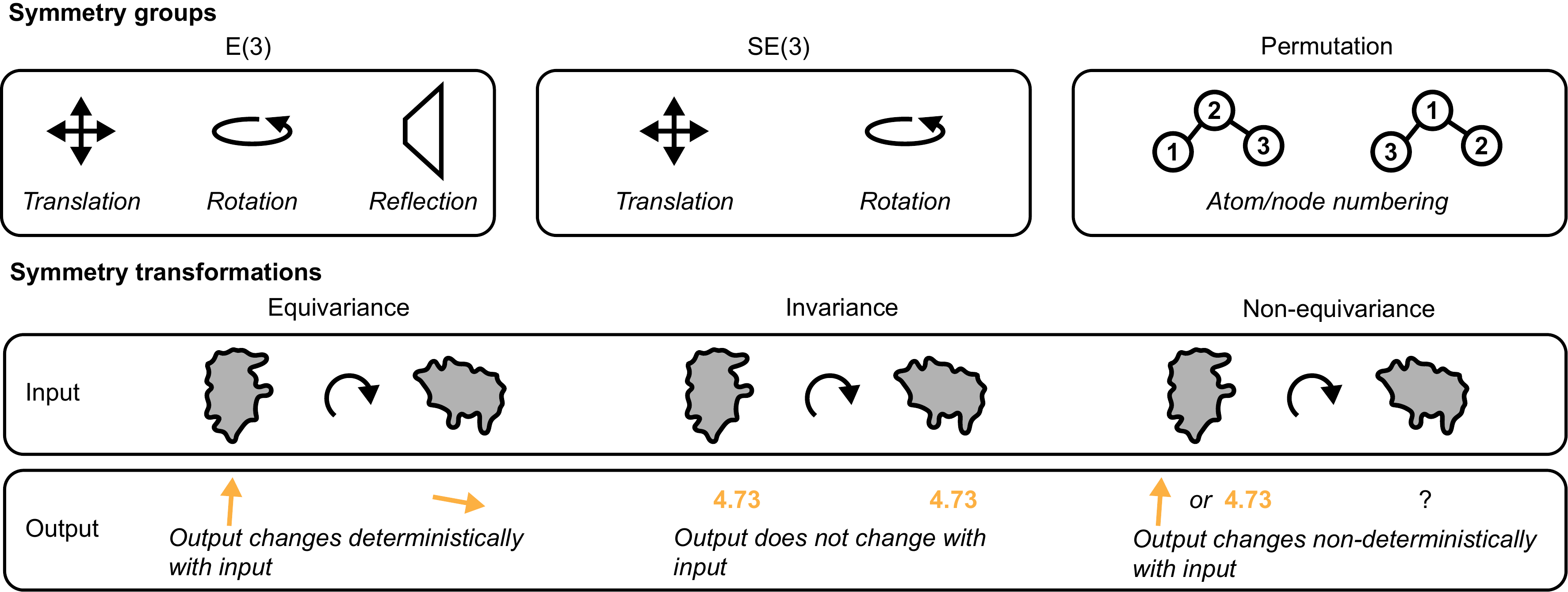}
\caption{Symmetry groups and transformations relevant for structure-based drug design. Symmetry transformations (lower panel) are shown exemplary as a rotation of a protein structure (gray) used as the input to a deep\-/learning model. The resulting effect on the output (orange) is shown both for the prediction of a vector (arrow) or a scalar (value: 4.73).}
\label{fig:equivariance}
\end{figure*}

Incorporating symmetry into the deep learning architecture to suit the input molecular representation and targeted property enables effective learning.~\cite{bronstein2017geometric} The most relevant symmetry groups of molecular systems include the Euclidean group E(3), the special Euclidean group SE(3), and the permutation group (Figure~\ref{fig:equivariance}).~\cite{atz2021geometric} Both E(3) and SE(3) cover transformations in a 3D coordinate system, including rotations and translations, but only E(3) covers reflections. Therefore, SE(3) becomes relevant when a neural network aims to learn different outputs for chiral inputs. Both symmetry groups are essential for learning in 3D. The permutation group is primarily related to the influence of node (\textit{i.e.} atom) numbering on neural network performance and is often incorporated \textit{via} permutation\-/invariant pooling operators (\textit{e.g.} sum, weighted-sum, max or mean). Concerning these three basic symmetry groups (E(3), SE(3) and permutation), the individual neural network layers $\mathcal{F}$ can transform the input $\mathcal{X}$ in an equivariant, invariant, or non-equivariant manner w.r.t the different symmetry properties.~\cite{atz2021geometric, bronstein2021geometric} \\

\begin{itemize}
    \item \textit{Equivariance.} $\mathcal{F}(\mathcal{X})$ is \textit{equivariant} to a transformation $\mathcal{T}$ if the transformation of input $\mathcal{X}$ commutes with the transformation of $\mathcal{F}(\mathcal{X})$ \textit{via} a transformation $\mathcal{T}'$ of the same symmetry group: $\mathcal{F}(\mathcal{T}(\mathcal{X})) = \mathcal{T}'(\mathcal{F}(\mathcal{X}))$. An example would be an E(3)\-/equivariant neural network that has access to the coordinate system (\textit{e.g.}, edge features that include relative coordinates) and predicts the dipole vector which rotates together with the input molecule. 
    
    \item \textit{Invariance.} Invariance is a special case of equivariance, where $\mathcal{F}(\mathcal{X})$ is invariant to $\mathcal{T}$ if $\mathcal{T}'$ is the trivial group action: $\mathcal{F}(\mathcal{T}(\mathcal{X})) = \mathcal{T}'(\mathcal{F}(\mathcal{X})) = \mathcal{F}(\mathcal{X})$. An example would be an E(3)\-/invariant neural network which has no access to the coordinate system, (\textit{e.g.}, edge features that only include pairwise distances) and predicts the same formation energy irrespective of how the input molecule is rotated, translated, or mirrored.  
    
    \item \textit{Non\-/equivariance.} $\mathcal{F}(\mathcal{X})$ is neither equivariant nor invariant to $\mathcal{T}$ when the transformation of the input $\mathcal{X}$ does not commute with the transformation of $\mathcal{F}(\mathcal{X})$: $\mathcal{F}(\mathcal{T}(\mathcal{X})) \neq \mathcal{T}'(\mathcal{F}(\mathcal{X}))$. An example would be an E(3)\-/non\-/equivariant neural network that has access to the coordinate system (\textit{e.g.}, explicit spatial properties such as voxels in a 3D grid) for which the output does change, but not deterministically, w.r.t to the alignment of the molecule. In such cases approximate invariance or equivariance can be learned through data augmentation.
\end{itemize}

Neural-network-based approaches featuring equivariance~\cite{weiler20183d}, invariance~\cite{schutt2018schnet}, and non\-/equivariance~\cite{jimenez2018k} (approximate invariance) to the E(3) or SE(3) symmetry groups are discussed below. 

\section{Molecular property prediction}
\label{sec:property_prediction}
This section discusses approaches that aim to predict a scalar quantity based on a macromolecular structure (potentially including a ligand), \textit{e.g.}, ligand-binding affinity prediction or docking pose scoring (see Figure~\ref{fig:classification_overview}, first row).

\subsection{Grid-based methods}
Several approaches represent the macromolecular structure on a 3D grid and employ convolutional neural networks (CNNs) to predict the property of interest. KDEEP estimates absolute binding affinities by representing the protein-ligand complex as a 3D grid in which each voxel is featurized with channels encoding a set of pharmacological properties separately for protein and ligand. Owing to the lack of rotational invariance of 3D\-/CNNs, \SI{90}{\degree} rotations of the input at training time were used for data augmentation.~\cite{jimenez2018k} Extending the approach of classical 3D\-/CNNs, 3D steerable CNNs can provide SE(3)\-/equivariant convolutions on grid-like data and have been used for the prediction of amino acid preference for the atomic environment and of the protein structural class. SE(3)\-/equivariance is achieved by using a linear combination of steerable kernels.~\cite{weiler20183d} 

% 3D surface
\subsection{Surface-based methods}
HoloProt, an approach for binding affinity and protein function prediction, encodes proteins across different length\-/scales by combining sequence\-/, surface\nobreakdash-, and structure\-/based graph representations. A surface\-/level graph uses nodes on the triangulated protein surface annotated with physicochemical and geometric features, whereas a structure\-/level graph based on amino acid residue nodes captures the 3D structure. A multi\-/level message\-/passing network aggregates the information from surface\-/ and structure\-/based representations, and combines it with the ligand graph to finally output the desired quantity (for binding affinity prediction).~\cite{somnath2021multi}

\begin{figure*}[ht!]
\centering
\includegraphics[width=\linewidth]{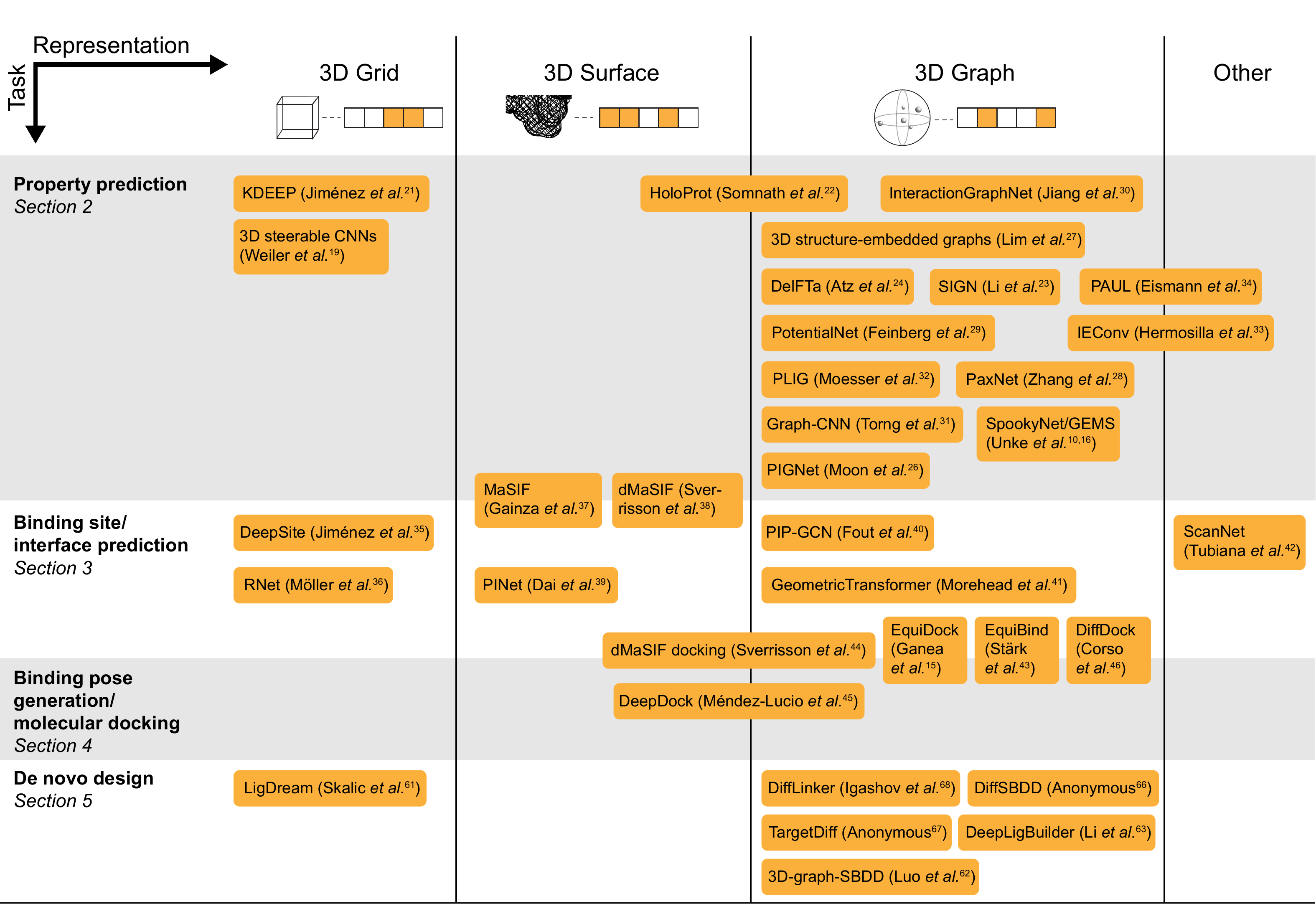}
\caption{Overview of geometric deep\-/learning methods for structure\-/based drug design discussed in this review. Methods are placed according to the task (rows) and macromolecular representation (columns).}
\label{fig:classification_overview}
\end{figure*}

% 3D graph
\subsection{Graph-based methods}
Various methods use 3D graphs to capture the structure of a macromolecule and combine it with ligand information, either \textit{via} separate ligand encoding or a macromolecule\-/ligand co\-/complex. By using 3D graphs instead of directly operating on the Cartesian atom coordinates, these approaches are often invariant to translation and rotation of the input structure. \\

The construction of 3D graphs differs among the various graph\-/based approaches. They either use an encoding of the node distances (atom\-/atom distance or residue\-/residue distance) as edge features, or different edge types (\textit{e.g.}, intra\-/ and intermolecular edges to differentiate between ligand and protein subgraphs), or construct edges in the molecular graph if the node distance falls below a pre-defined threshold. These methods for graph construction are not mutually exclusive, and combinations between them exist. \\

As an example for directly using node distances, SIGN predicts protein\-/ligand binding affinities by using iterative interaction layers with either angle or distance awareness, to incorporate knowledge of spatial orientation during the message\-/passing steps.~\cite{li2021structure} Another E(3)\-/invariant architecture, the 3D message\-/passing neural network DelFTa~\cite{atz2022delta}, trained on a data set of quantum\-/mechanical reference calculations~\cite{isert2022qmugs}, uses Fourier\-/encoded atom distances to predict Wiberg bond orders of macromolecular molecules.~\cite{atz2022delta} Based on the E(3)\-/equivariant 3D message\-/passing neural network architecture SpookyNet~\cite{unke2021spookynet}, GEMS trains a machine learning force field on molecular fragments to obtain close-to-DFT accuracy for protein structures. By incorporating long\-/range interactions learned from top\-/down generated fragments, greater\-/than\-/expected flexibility of protein structures was discovered.~\cite{unke2022accurate} \\

By combining direct distance encoding and different edge types, PIGNet~\cite{moon2022pignet} aims to predict binding affinities. To this end, PIGNet uses physics\-/informed pairwise interactions modeled with a gate\-/augmented graph attention network~\cite{moon2022pignet} In another approach, Lim \textit{et al.}~\cite{lim2019predicting} classified ligands as either active or inactive against a given target by treating covalent and non\-/covalent interactions separately, while using distances for the description of non\-/covalent interactions. By subtracting the graph features obtained from the protein and ligand separately from those of the respective complex, the relevance of intermolecular interactions can be perceived by the network.~\cite{lim2019predicting} Tackling the binding affinity prediction problem, PaxNet differentiates between local and non\-/local interactions with separate (but connected) message-passing schemes for each type of interaction. By using angle information only for local interactions and focusing on distances for non\-/local interactions, the computational cost of geometric deep learning operations can be minimized.~\cite{zhang2022efficient} \\

Examples of using different edge types include PotentialNet~\cite{feinberg2018potentialnet} and InteractionGraphNet~\cite{jiang2021interactiongraphnet} for binding affinity prediction, differentiating between covalent and non\-/covalent, respectively intra\-/ and intermolecular graph convolutions. In another approach, Torng \textit{et al.}~\cite{torng2019graph} used an unsupervised graph\-/autoencoder to generate representative binding pocket representations followed by protein\-/level graph convolutions based on a Euclidean distance cutoff for active/inactive classification of a protein-ligand pair.~\cite{torng2019graph} \\

Departing from the direct use of the protein structure in the 3D graph, the recently introduced "Protein-Ligand Interaction Graphs" (PLIGs) incorporate information about the protein environment directly into the node features of the ligand graph, thereby reducing the size and complexity of the resulting graph structure.~\cite{moesser2022protein}

% Other
\subsection{Other methods}
In addition to grids, surfaces, and graphs, various types of data have been used to model macromolecular structures. For example, Hermosilla \textit{et al.}~\cite{hermosilla2020intrinsic} approached the problem of protein fold and reaction classification as a learning problem on 3D point clouds, introducing an E(3)\-/invariant convolution operator that considers extrinsic (Euclidean) and intrinsic distances (covalent-only or covalent and non\-/covalent bond hop distance).~\cite{hermosilla2020intrinsic} A network architecture called PAUL predicts the root\-/mean\-/square\-/deviation of protein structure from an experimental structure directly from the 3D coordinates of atoms by using SE(3)\-/equivariant convolution filters.~\cite{eismann2021hierarchical}

\section{Binding site/interface prediction}
\label{sec:binding_site}
This section discusses approaches that aim to predict the parts of a macromolecular structure that can act as a binding site for small, drug\-/like ligands, or as an interaction interface for other macromolecules (see Figure~\ref{fig:classification_overview}, second row). Binding pose-generating methods that implicitly identify binding sites are discussed in Section~\ref{sec:docking}.

% 3D grids
\subsection{Grid-based methods}
DeepSite is an early approach that represents a protein using a regular 3D grid with voxels characterized by pharmacophoric features of nearby atom types. Using a sliding subgrid approach, the network outputs the probability that this sub-grid is close to a druggable binding site.~\cite{jimenez2017deepsite} RNet extended this approach to predicting ligand-binding sites of ribonucleic acid (RNA) structures.~\cite{moller2022translating}

% 3D surface
\subsection{Surface-based methods}
MaSIF~\cite{gainza2020deciphering} (molecular surface interaction fingerprinting) and its differentiable analogue dMaSIF~\cite{sverrisson2021fast} use macromolecular surface representations for binding site prediction, and can also classify, \textit{e.g.}, pocket function. The surface-based approach describes individual points on the protein surface in geodesic space, such that distances between points correspond to the length of the path between them along the surface, rather than to the Euclidean distance. In a three\-/step approach, the surface is decomposed into individual patches. Points within each patch are featurized with geometric and chemical properties. Geodesic convolutions transform these features into a numerical vector for downstream tasks. The first two steps required expensive pre-computation in the original implementation~\cite{gainza2020deciphering}, whereas dMaSIF is end\-/to\-/end differentiable and operates directly on atom types and coordinates.~\cite{sverrisson2021fast} Another approach, PINet, uses a physics\-/inspired geometric deep learning network to identify interface regions between two interacting proteins by learning the complementarity of surface shape and physicochemical properties. Owing to the lack of rotational invariance in this network architecture, data augmentation using random rotations of the input structures is required.~\cite{dai2021protein}

% 3D graph
\subsection{Graph-based methods}
Networks operating on 3D graph representations of molecular structures have been widely applied to binding sites and interface prediction. Examples of such approaches are rototranslational\-/invariant methods that use edge features (including distances and angles) to infuse the model with geometric understanding, followed by the prediction of pairwise residue\-/level interaction potential using either spatial graph convolutions~\cite{fout2017protein} or a graph transformer~\cite{morehead2021geometric}.

\subsection{Other approaches}
ScanNet uses an E(3)\-/invariant geometric deep learning model by applying structure\-/based linear Gaussian kernel filters to predict protein\-/protein and protein\-/antibody binding sites.~\cite{tubiana2022scannet}

\section{Binding pose generation/molecular docking}
\label{sec:docking}
This section focuses on methods for docking pose generation, \textit{i.e.}, the generation of a suitable binding conformation between for either a small-molecule ligand and its macromolecular receptor, or between two macromolecular structures (see Figure~\ref{fig:classification_overview}, third row).

\begin{figure*}[th!]
\centering
\includegraphics[width=\linewidth]{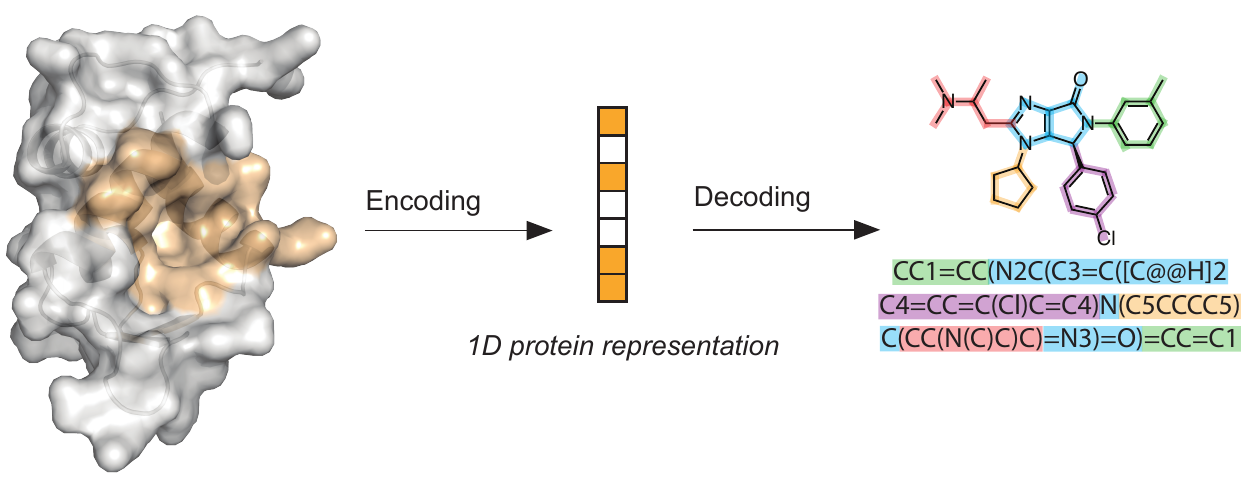}
\caption{Structure-based \textit{de novo} design with machine learning. A protein, or only its ligand-binding site, is processed using an encoder neural network, which transforms the protein representation into a 1D latent space. This latent space can then be transformed into SMILES-strings by a decoder neural network such as a chemical language model. The illustrated molecule DF-1 was designed using a novel, yet unpublished geometric deep-learning method to target MDM2 (\textit{cf.} Section \ref{section:df}). The active binding site is colored on the protein surface. The 1D protein representation corresponds to the feature vector of the latent space.}
\label{fig:denovo}
\end{figure*}

\subsection{Graph-based and hybrid methods}
EquiDock uses an SE(3)-equivariant message passing neural network combined with optimal transport to predict the binding pose between two protein molecules in a rigid-body, blind-docking fashion (which also entails detecting a suitable binding site). The network predicts a rotation matrix and a translation vector to move one protein structure into the binding pose while keeping the second protein fixed, guaranteeing that the resulting docking pose is invariant to the initial orientation and positioning of both binding partners.~\cite{ganea2021independent} EquiBind extends this approach to the docking of flexible small-molecule ligand molecules to protein structures, incorporating changes in the torsion angles of rotatable bonds starting from a randomly generated conformer.~\cite{stark2022equibind} Building on dMaSIF~\cite{sverrisson2021fast}, another approach for rigid-body docking of protein complexes combines SE(3)\-/equivariant graph neural networks with surface fingerprints for atomic point clouds to estimate the surface shape complementarity of the two binding partners.~\cite{sverrisson2022physics} In contrast to EquiDock~\cite{ganea2021independent}, this approach generates and ranks multiple binding poses for each protein-protein pair.~\cite{sverrisson2022physics} DeepDock constitutes a geometric deep-learning approach for predicting small\-/molecule binding poses by representing the binding site surface as a polygon mesh and the ligand as a molecular graph, and predicting a probability distribution over pairwise node distances between the ligand and protein.~\cite{mendez2021geometric} DiffDock uses a diffusion\-/based generative model for molecular docking. The approach generates a tuneable number of ligand poses in a two\-/step process: First, a scoring model uses a reverse diffusion process that transforms random initial ligand poses into predicted poses by translation, rotation, and torsion angle changes. Second, a confidence model predicts a binary label indicating whether a generated ligand pose is below the 2 Å root\-/mean\-/square\-/distance threshold commonly used to evaluate binding pose accuracy. While the scoring model employs a residue-level 3D graph of the protein, the confidence model uses a full-atomistic 3D graph. The translation and rotation outputs of the scoring model are SE(3)\-/equivariant, whereas the torsion angle outputs and confidence model predictions are SE(3)\-/invariant. The approach substantially outperformed existing classical and deep learning\-/based approaches on a common docking benchmark.~\cite{corso2022diffdock}

\section{De novo design} 
\label{sec:denovo}
\textit{De novo} design methods aim to generate new molecular structures with desired biological and physical properties from scratch (see Figure~\ref{fig:classification_overview}, fourth row).~\cite{schneider2005computer} The concept of deriving features from a protein binding site tailored for the use in automated \textit{de novo} design dates back more than thirty years.~\cite{danziger1989automated} Early structure-based \textit{de novo} design approaches used to generate desired molecular structures iteratively by using either single atoms or fragments.~\cite{schneider2000novo} Prominent algorithms included (i) linking~\cite{bohm1992computer} (\textit{i.e.} placing building blocks at key interaction sites of the receptor and linking them), (ii) growing~\cite{rotstein1993groupbuild} (\textit{i.e.} starting with a single building block at one of the key interaction sites of the receptor and growing to a complete ligand), and (iii) lattice-based design~\cite{lewis1992automated} (\textit{i.e.} starting from an atomic lattice in the receptor binding site consisting of sp$^3$ carbon atoms and replacing them until a complete ligand is formed).

\subsection{Chemical language models}
More recently, deep learning has been used for \textit{de novo} molecular design and has found various applications in medicinal chemistry and chemical biology.~\cite{schneider2016novo} Currently, the most prevalent and successful deep learning models for \textit{de novo} drug design, so called chemical language models (CLMs), learn on string-based molecular representations (\textit{e.g.}, SMILES strings).~\cite{segler2018generating} Ligand-based \textit{de novo} design with CLMs has led to the successful generation of molecules with desired physicochemical and biological properties.~\cite{merk2018novo, merk2018tuning, grisoni2021combining, yuan2017chemspacemim} In this context, data augmentation based on non-canonical SMILES string enumeration~\cite{arus2019randomized} and bidirectional learning~\cite{grisoni2020bidirectional} has been shown to considerably improve the quality of the chemical language learned by CLMs. Such ligand-based deep generative methods have been extended to structure-based approaches that incorporate explicit information of the targeted protein (Figure~\ref{fig:denovo}). Convolutional neural networks using 3D-grid-based representation of the binding site of the protein as input have been proposed to learn a latent space that is then decoded to sequences (\textit{i.e.} SMILES\-/string) using a CLM.~\cite{skalic2019shape} 

\subsection{Graph-based methods}
Methods have been proposed that generate 3D structures of potential ligand molecules directly from the 3D structure of the macromolecular binding site. Ligands are constructed within the binding site in the form of a 3D graph.~\cite{luo20213d, li2021structure_based} These models sample atoms sequentially from a learned distribution and have been shown to be applicable to a variety of molecular properties.~\cite{gebauer2022inverse} Recent work has introduced E(3)\-/equivariant diffusion models that enable the generation of molecules as 3D graphs by learning to denoise a normally distributed set of points.~\cite{hoogeboom2022equivariant} This process has been extended to molecule generation from scratch within the binding site of macromolecules, as implemented in DiffSBDD~\cite{anonymous2023structurebased} and TargetDiff~\cite{anonymous2023d}. DiffLinker generates suitable linkers that connect fragments placed in the binding pocket~\cite{igashov2022equivariant}. Although these 3D graph\-/based \textit{de novo} design models can construct a large fraction of novel molecules, their practical applications remain to be explored.

\section{Outlook}
Based on the success of these pioneering applications, we expect future work to extend equivariant neural networks and physics-inspired approaches to structure-based drug design. Previous studies have indicated that incorporating certain aspects of physics and symmetry into a model tends to increase the accuracy, generalizability, and interpretability of the predictions.~\cite{moon2022pignet, batatia2022mace, batzner20223} We further expect that geometric deep learning research for structure\-/based drug design will follow trends in the pharmaceutical industry. These trends include a growing interest in protein-protein interaction inhibitors~\cite{cheng2020design}, induced proximity approaches such as molecular glues~\cite{schreiber2021rise} and proteolysis targeting chimeras (PROTACs)~\cite{li2022protacs}, as well as RNA-targeting therapeutics~\cite{salton2016small}. \\

Novel applications in the field of structure\-/based binding affinity prediction will have to address the points of criticism directed toward existing methods. Recent work has shown that many deep learning architectures trained on the PDBbind~\cite{wang2004pdbbind} data set merely memorize the training data rather than learning a meaningful mapping between protein\-/ligand structure and binding affinity, contributing to poor generalization performance in some cases. The use of protein-ligand complexes for model development often leads to similar performance compared to the use of ligand\-/only or protein\-/only descriptors.~\cite{volkov2022frustration} Future work in this area will likely benefit from suitable benchmarking data sets~\cite{parks2020d3r,gaieb2018d3r}, and guidelines for building such data sets have been proposed.~\cite{hahn2021best} A promising example for such a data set is the recently released collection of binding affinities and X-ray co-crystal structures of PDE10A inhibitors.~\cite{tosstorff2022high} Included train-test splits mirroring different real-world lead optimization scenarios can be used to assess generalization performance. \\

3D-aware models, such as normalizing flow-based approaches, may appear at the forefront of future generative modeling studies.~\cite{hoogeboom2022equivariant} To comprehensively evaluate the utility of emerging new models in a real\-/world drug design context, experimental validation of the proposed molecular structures is paramount. Since not all computational groups working in this field will have the expertise, equipment, or desire to perform the required synthesis and experimental testing, collaborations with experimentalists will be highly valuable.

\section{Acknowledgements}
This work was financially supported by the Swiss National Science Foundation (grant no. 205321\_182176). C.I. acknowledges support from the Scholarship Fund of the Swiss Chemical Industry. \\ \\

\section{Competing interest}
G.S. declares a potential financial conflict of interest as co-founder of inSili.com LLC, Zurich, and in his role as a scientific consultant to the pharmaceutical industry.

\section{Structure-based de novo design of DF-1}
The ligand DF-1 displayed in Figures \ref{fig:representation_overview} and \ref{fig:denovo} was designed using a new geometric deep-learning method to target MDM2 (K. Atz, G. Schneider, unpublished). DF-1 was docked into the active binding site of the human MDM2 protein (PDB-ID 4JRG) using GOLD software~\cite{verdonk2003improved}. The highest ranking docking pose is shown in Figure~\ref{fig:representation_overview}. DF-1 has an Euclidean distance of 0.48 (using ECFP4~\cite{rogers2010extended}) to the closest molecule (ChEMBL3653229) in the ChEMBL database~\cite{mendez2019chembl} (Version 29), and atomic- and graph-scaffolds that are not present in this database. 
\label{section:df}

%%%%%%%%%
%%%%%%%%%
%%%%%%%%%
%%%%%%%%%

\printbibliography

@article{atz2021geometric,
  title={Geometric deep learning on molecular representations},
  author={Atz, Kenneth and Grisoni, Francesca and Schneider, Gisbert},
  journal={Nat. Mach. Intell.},
  pages={1023--1032},
  volume = {3},
  year={2021},
  publisher={Nature Publishing Group},
}

@article{yuan2017chemspacemim,
    author = {Yuan, William and Jiang, Dadi and Nambiar, Dhanya K. and Liew, Lydia P. and Hay, Michael P. and Bloomstein, Joshua and Lu, Peter and Turner, Brandon and Le, Quynh-Thu and Tibshirani, Robert and Khatri, Purvesh and Moloney, Mark G. and Koong, Albert C.},
    title = {Chemical Space Mimicry for Drug Discovery},
    journal = {J. Chem. Inf. Model.},
    volume = {57},
    number = {4},
    pages = {875-882},
    year = {2017},
}

@article{tosstorff2022high,
  title={A high quality, industrial data set for binding affinity prediction: performance comparison in different early drug discovery scenarios},
  author={Tosstorff, Andreas and Rudolph, Markus G and Cole, Jason C and Reutlinger, Michael and Kramer, Christian and Schaffhauser, Herv{\'e} and Nilly, Agn{\`e}s and Flohr, Alexander and Kuhn, Bernd},
  journal={J. Comput. Aided Mol. Des.},
  pages={1--13},
  year={2022},
  publisher={Springer},
}

@article{arus2019randomized,
  title={Randomized SMILES strings improve the quality of molecular generative models},
  author={Ar{\'u}s-Pous, Josep and Johansson, Simon Viet and Prykhodko, Oleksii and Bjerrum, Esben Jannik and Tyrchan, Christian and Reymond, Jean-Louis and Chen, Hongming and Engkvist, Ola},
  journal={J. Cheminformatics},
  volume={11},
  number={1},
  pages={1--13},
  year={2019},
  publisher={BioMed Central}
}

@article{grisoni2020bidirectional,
  title={Bidirectional molecule generation with recurrent neural networks},
  author={Grisoni, Francesca and Moret, Michael and Lingwood, Robin and Schneider, Gisbert},
  journal={J. Chem. Inf. Model.},
  volume={60},
  number={3},
  pages={1175--1183},
  year={2020},
  publisher={ACS Publications}
}

@article{danziger1989automated,
  title={Automated site-directed drug design: a general algorithm for knowledge acquisition about hydrogen-bonding regions at protein surfaces},
  author={Danziger, DJ and Dean, PM},
  journal={Proc. Royal Soc. B .},
  volume={236},
  number={1283},
  pages={101--113},
  year={1989},
  publisher={The Royal Society London}
}

@article{schneider2000novo,
  title={De novo design of molecular architectures by evolutionary assembly of drug-derived building blocks},
  author={Schneider, Gisbert and Lee, Man-Ling and Stahl, Martin and Schneider, Petra},
  journal={J. Comput. Aided Mol. Des.},
  volume={14},
  number={5},
  pages={487--494},
  year={2000},
  publisher={Springer}
}

@article{rotstein1993groupbuild,
  title={GroupBuild: a fragment-based method for de novo drug design},
  author={Rotstein, Sergio H and Murcko, Mark A},
  journal={J. Med. Chem.},
  volume={36},
  number={12},
  pages={1700--1710},
  year={1993},
  publisher={ACS Publications}
}

@article{bohm1992computer,
  title={The computer program LUDI: a new method for the de novo design of enzyme inhibitors},
  author={B{\"o}hm, Hans-Joachim},
  journal={J. Comput. Aided Mol. Des.},
  volume={6},
  number={1},
  pages={61--78},
  year={1992},
  publisher={Springer}
}

@article{lewis1992automated,
  title={Automated site-directed drug design using molecular lattices},
  author={Lewis, Richard A and Roe, Diana C and Huang, Conrad and Ferrin, Thomas E and Langridge, Robert and Kuntz, Irwin D},
  journal={J. Mol. Graph.},
  volume={10},
  number={2},
  pages={66--78},
  year={1992},
  publisher={Elsevier}
}

@article{schneider2005computer,
  title={Computer-based de novo design of drug-like molecules},
  author={Schneider, Gisbert and Fechner, Uli},
  journal={Nat. Rev. Drug Discovery},
  volume={4},
  number={8},
  pages={649--663},
  year={2005},
  publisher={Nature Publishing Group}
}

@article{nippa2022enabling,
  title={Enabling late-stage drug diversification by high-throughput experimentation with geometric deep learning},
  author={Nippa, David F. and Atz, Kenneth and Hohler, Remo and M{\"u}ller, Alex T. and Marx, Andreas and Bartelmus, Christian and Wuitschik, Georg and Marzuoli, Irene and Jost, Vera and Wolfard, Jens and Binder, Martin and Stepan, Antonia F. and Konrad, David B. and Grether, Uwe and Martin, Rainer E. and Schneider, Gisbert},
  journal={ChemRxiv preprint 10.26434/chemrxiv-2022-gkxm6},
  year={2022}
}

@article{luo20213d,
  title={A 3D generative model for structure-based drug design},
  author={Luo, Shitong and Guan, Jiaqi and Ma, Jianzhu and Peng, Jian},
  journal={Advances in Neural Information Processing Systems (NeurIPS)},
  volume={34},
  pages={6229--6239},
  year={2021}
}

@article{li2021structure_based,
  title={Structure-based de novo drug design using 3D deep generative models},
  author={Li, Yibo and Pei, Jianfeng and Lai, Luhua},
  journal={Chem. Sci.},
  volume={12},
  number={41},
  pages={13664--13675},
  year={2021},
  publisher={Royal Society of Chemistry}
}

@article{verdonk2003improved,
  title={Improved protein--ligand docking using GOLD},
  author={Verdonk, Marcel L and Cole, Jason C and Hartshorn, Michael J and Murray, Christopher W and Taylor, Richard D},
  journal={Proteins: Struct., Funct., Bioinf.},
  volume={52},
  number={4},
  pages={609--623},
  year={2003},
  publisher={Wiley Online Library}
}

@article{rogers2010extended,
  title={Extended-connectivity fingerprints},
  author={Rogers, David and Hahn, Mathew},
  journal={J. Chem. Inf. Model.},
  volume={50},
  number={5},
  pages={742--754},
  year={2010},
  publisher={ACS Publications}
}

@article{mendez2019chembl,
  title={{ChEMBL}: Towards direct deposition of bioassay data},
  author={Mendez, David and Gaulton, Anna and Bento, A Patr{\'\i}cia and Chambers, Jon and De Veij, Marleen and F{\'e}lix, Eloy and Magari{\~n}os, Mar{\'\i}a Paula and Mosquera, Juan F and Mutowo, Prudence and Nowotka, Micha{\l} and others},
  journal={Nucleic Acids Res.},
  volume={47},
  number={D1},
  pages={D930--D940},
  year={2019},
  publisher={Oxford University Press}
}

@article{li2021structure,
  title={Structure-aware interactive graph neural networks for the prediction of protein-ligand binding affinity},
  author={Li, Shuangli and Zhou, Jingbo and Xu, Tong and Huang, Liang and Wang, Fan and Xiong, Haoyi and Huang, Weili and Dou, Dejing and Xiong, Hui},
  journal={Proceedings of the 27th ACM SIGKDD Conference on Knowledge Discovery \& Data Mining},
  pages={975--985},
  year={2021}
}

@article{gebauer2022inverse,
  title={{Inverse design of 3D molecular structures with conditional generative neural networks}},
  author={Gebauer, Niklas WA and Gastegger, Michael and Hessmann, Stefaan SP and M{\"u}ller, Klaus-Robert and Sch{\"u}tt, Kristof T},
  journal={Nat. Commun.},
  volume={13},
  pages={973},
  year={2022},
  publisher={Nature Publishing Group}
}

@article{sledz2018protein,
  title={Protein structure-based drug design: from docking to molecular dynamics},
  author={{\'S}led{\'z}, Pawe{\l} and Caflisch, Amedeo},
  journal={Curr. Opin. Struct. Biol.},
  volume={48},
  pages={93--102},
  year={2018},
  publisher={Elsevier}
}

@article{schutt2018schnet,
  title={{SchNet -- a deep learning architecture for molecules and materials}},
  author={Sch{\"u}tt, Kristof T and Sauceda, Huziel E and Kindermans, P-J and Tkatchenko, Alexandre and M{\"u}ller, K-R},
  journal={J. Chem. Phys.},
  volume={148},
  number={24},
  pages={241722},
  year={2018},
  publisher={AIP Publishing LLC}
}

@article{bleicher2003hit,
  title={Hit and lead generation: beyond high-throughput screening},
  author={Bleicher, Konrad H and B{\"o}hm, Hans-Joachim and M{\"u}ller, Klaus and Alanine, Alexander I},
  journal={Nat. Rev. Drug Discov.},
  volume={2},
  number={5},
  pages={369--378},
  year={2003},
  publisher={Nature Publishing Group}
}

@article{unke2021spookynet,
  title={{SpookyNet: Learning force fields with electronic degrees of freedom and nonlocal effects}},
  author={Unke, Oliver T and Chmiela, Stefan and Gastegger, Michael and Sch{\"u}tt, Kristof T and Sauceda, Huziel E and M{\"u}ller, Klaus-Robert},
  journal={Nat. Commun.},
  volume={12},
  pages={7273},
  year={2021},
  publisher={Nature Publishing Group},

}

@article{bissantz2010medicinal,
  title={A medicinal chemist’s guide to molecular interactions},
  author={Bissantz, Caterina and Kuhn, Bernd and Stahl, Martin},
  journal={J. Med. Chem.},
  volume={53},
  number={14},
  pages={5061--5084},
  year={2010},
  publisher={ACS Publications},
}

@article{bronstein2021geometric,
  title={Geometric deep learning: Grids, groups, graphs, geodesics, and gauges},
  author={Bronstein, Michael M and Bruna, Joan and Cohen, Taco and Veli{\v{c}}kovi{\'c}, Petar},
  journal={arXiv preprint arXiv:2104.13478},
  year={2021},
  addendum={\newline \textbf{* This work provides an introduction to the terminology of geometric deep learning as a unification of machine learning for different data structures and neural network architectures from the perspective of symmetry and invariance.}},
}

@article{stark2022equibind,
  title={{EquiBind: Geometric deep learning for drug binding structure prediction}},
  author={St{\"a}rk, Hannes and Ganea, Octavian and Pattanaik, Lagnajit and Barzilay, Regina and Jaakkola, Tommi},
  journal={International Conference on Machine Learning (ICML)},
  volume={39},
  pages={20503--20521},
  year={2022},
  organization={PMLR}
}

@article{merk2018tuning,
  title={Tuning artificial intelligence on the de novo design of natural-product-inspired retinoid X receptor modulators},
  author={Merk, Daniel and Grisoni, Francesca and Friedrich, Lukas and Schneider, Gisbert},
  journal={Commun. Chem.},
  volume={1},
  number={1},
  pages={1--9},
  year={2018},
  publisher={Nature Publishing Group}
}

@article{skalic2019shape,
  title={Shape-based generative modeling for de novo drug design},
  author={Skalic, Miha and Jim{\'e}nez, Jos{\'e} and Sabbadin, Davide and De Fabritiis, Gianni},
  journal={J. Chem. Inf. Model.},
  volume={59},
  number={3},
  pages={1205--1214},
  year={2019},
  publisher={ACS Publications}
}

@article{unke2022accurate,
  title={{Accurate Machine Learned Quantum-Mechanical Force Fields for Biomolecular Simulations}},
  author={Unke, Oliver T and St{\"o}hr, Martin and Ganscha, Stefan and Unterthiner, Thomas and Maennel, Hartmut and Kashubin, Sergii and Ahlin, Daniel and Gastegger, Michael and Sandonas, Leonardo Medrano and Tkatchenko, Alexandre and others},
  journal={arXiv preprint arXiv:2205.08306},
  year={2022},
  addendum={\newline\textbf{** GEMS introduces an approach to extend machine-learning force fields to large macromolecular structures by training on molecular fragments. Using a physically motivated architecture, selected protein dynamics and protein-protein interactions are modelled at close to ab initio accuracy.}},
}

@article{isert2022qmugs,
  title={{QMugs, quantum mechanical properties of drug-like molecules}},
  author={Isert, Clemens and Atz, Kenneth and Jim{\'e}nez-Luna, Jos{\'e} and Schneider, Gisbert},
  journal={Sci. Data},
  volume={9},
  pages={273},
  year={2022},
  publisher={Nature Publishing Group},
}

@article{christensen2021orbnet,
  title={{OrbNet Denali: A machine learning potential for biological and organic chemistry with semi-empirical cost and DFT accuracy}},
  author={Christensen, Anders S and Sirumalla, Sai Krishna and Qiao, Zhuoran and O’Connor, Michael B and Smith, Daniel GA and Ding, Feizhi and Bygrave, Peter J and Anandkumar, Animashree and Welborn, Matthew and Manby, Frederick R and others},
  journal={J. Chem. Phys.},
  volume={155},
  number={20},
  pages={204103},
  year={2021},
  publisher={AIP Publishing LLC},
  addendum={\newline\textbf{* OrbNet Denali introduces a message-passing mechanism for graph neural networks that uses symmetry-adapted atomic orbital features from a low-cost QM calculation as input features.}},

}

@article{unke2021se,
  title={{SE(3)-equivariant prediction of molecular wavefunctions and electronic densities}},
  author={Unke, Oliver and Bogojeski, Mihail and Gastegger, Michael and Geiger, Mario and Smidt, Tess and M{\"u}ller, Klaus-Robert},
  journal={Advances in Neural Information Processing Systems (NeurIPS)},
  volume={34},
  pages={14434--14447},
  year={2021}
}

@article{satorras2021n,
  title={E(n) equivariant graph neural networks},
  author={Satorras, V{\i}ctor Garcia and Hoogeboom, Emiel and Welling, Max},
  journal={International Conference on Machine Learning (ICML)},
  volume={38},
  pages={9323--9332},
  year={2021},
}

@article{batatia2022mace,
  title={Mace: Higher order equivariant message passing neural networks for fast and accurate force fields},
  author={Batatia, Ilyes and Kov{\'a}cs, D{\'a}vid P{\'e}ter and Simm, Gregor NC and Ortner, Christoph and Cs{\'a}nyi, G{\'a}bor},
  journal={arXiv preprint arXiv:2206.07697},
  year={2022}
}

@article{batzner20223,
  title={{E(3)-equivariant graph neural networks for data-efficient and accurate interatomic potentials}},
  author={Batzner, Simon and Musaelian, Albert and Sun, Lixin and Geiger, Mario and Mailoa, Jonathan P and Kornbluth, Mordechai and Molinari, Nicola and Smidt, Tess E and Kozinsky, Boris},
  journal={Nat. Commun.},
  volume={13},
  pages={2453},
  year={2022},
  publisher={Nature Publishing Group}
}

@article{bronstein2017geometric,
  title={{Geometric deep learning: going beyond Euclidean data}},
  author={Bronstein, Michael M and Bruna, Joan and LeCun, Yann and Szlam, Arthur and Vandergheynst, Pierre},
  journal={IEEE Signal Process. Mag.},
  volume={34},
  number={4},
  pages={18--42},
  year={2017},
  publisher={IEEE}
}

@article{atz2022delta,
  title={{$\Delta$-Quantum machine-learning for medicinal chemistry}},
  author={Atz, Kenneth and Isert, Clemens and B{\"o}cker, Markus NA and Jim{\'e}nez-Luna, Jos{\'e} and Schneider, Gisbert},
  journal={Phys. Chem. Chem. Phys.},
  volume={24},
  number={18},
  pages={10775--10783},
  year={2022},
  publisher={Royal Society of Chemistry}
}

@article{segler2018generating,
  title={{Generating focused molecule libraries for drug discovery with recurrent neural networks}},
  author={Segler, Marwin HS and Kogej, Thierry and Tyrchan, Christian and Waller, Mark P},
  journal={ACS Cent. Sci.},
  volume={4},
  number={1},
  pages={120--131},
  year={2018},
  publisher={ACS Publications},
}

@article{grisoni2021combining,
  title={{Combining generative artificial intelligence and on-chip synthesis for de novo drug design}},
  author={Grisoni, Francesca and Huisman, Berend JH and Button, Alexander L and Moret, Michael and Atz, Kenneth and Merk, Daniel and Schneider, Gisbert},
  journal={Sci. Adv.},
  volume={7},
  number={24},
  year={2021},
  pages={eabg3338},
  publisher={American Association for the Advancement of Science}
}

@article{merk2018novo,
  title={{De novo design of bioactive small molecules by artificial intelligence}},
  author={Merk, Daniel and Friedrich, Lukas and Grisoni, Francesca and Schneider, Gisbert},
  journal={Mol. Inform.},
  volume={37},
  number={1-2},
  pages={1700153},
  year={2018},
  publisher={Wiley Online Library}
}

@article{ganea2021independent,
  title={Independent SE (3)-Equivariant Models for End-to-End Rigid Protein Docking},
  author={Ganea, Octavian-Eugen and Huang, Xinyuan and Bunne, Charlotte and Bian, Yatao and Barzilay, Regina and Jaakkola, Tommi S and Krause, Andreas},
  journal={International Conference on Learning Representations (ICML)},
  volume={38},
  year={2021}
}

@article{schneider2016novo,
  title={{De novo design at the edge of chaos: Miniperspective}},
  author={Schneider, Petra and Schneider, Gisbert},
  journal={J. Med. Chem.},
  volume={59},
  number={9},
  pages={4077--4086},
  year={2016},
  publisher={ACS Publications}
}

@incollection{atz2023machine,
  title={Machine Learning and Computational Chemistry for the Endocannabinoid System},
  author={Atz, Kenneth and Guba, Wolfgang and Grether, Uwe and Schneider, Gisbert},
  booktitle={Endocannabinoid Signaling},
  pages={477--493},
  year={2022},
  publisher={Springer}
}

@article{sadybekov2022synthon,
  title={{Synthon-based ligand discovery in virtual libraries of over 11 billion compounds}},
  author={Sadybekov, Arman A and Sadybekov, Anastasiia V and Liu, Yongfeng and Iliopoulos-Tsoutsouvas, Christos and Huang, Xi-Ping and Pickett, Julie and Houser, Blake and Patel, Nilkanth and Tran, Ngan K and Tong, Fei and others},
  journal={Nature},
  volume={601},
  number={7893},
  pages={452--459},
  year={2022},
  publisher={Nature Publishing Group}
}

@article{jimenez2018k,
  title={{KDEEP: protein--ligand absolute binding affinity prediction via 3D-convolutional neural networks}},
  author={Jim{\'e}nez, Jos{\'e} and Skalic, Miha and Martinez-Rosell, Gerard and De Fabritiis, Gianni},
  journal={J. Chem. Inf. Model.},
  volume={58},
  number={2},
  pages={287--296},
  year={2018},
  publisher={ACS Publications}
}

@article{weiler20183d,
  title={{3D steerable CNNs: Learning rotationally equivariant features in volumetric data}},
  author={Weiler, Maurice and Geiger, Mario and Welling, Max and Boomsma, Wouter and Cohen, Taco S},
  journal={Advances in Neural Information Processing Systems (NeurIPS)},
  volume={31},
  year={2018}
}

@article{somnath2021multi,
  title={{Multi-scale representation learning on proteins}},
  author={Somnath, Vignesh Ram and Bunne, Charlotte and Krause, Andreas},
  journal={Advances in Neural Information Processing Systems (NeurIPS)},
  volume={34},
  pages={25244--25255},
  year={2021}, 
  addendum={\newline \textbf{* HoloProt combines different molecular representations to aggregate information from multiple length scales and predicts binding affinity and protein function.}},
}

@article{torng2019graph,
  title={{Graph convolutional neural networks for predicting drug-target interactions}},
  author={Torng, Wen and Altman, Russ B},
  journal={J. Chem. Inf. Model.},
  volume={59},
  number={10},
  pages={4131--4149},
  year={2019},
  publisher={ACS Publications}
}

@article{feinberg2018potentialnet,
  title={{PotentialNet for molecular property prediction}},
  author={Feinberg, Evan N and Sur, Debnil and Wu, Zhenqin and Husic, Brooke E and Mai, Huanghao and Li, Yang and Sun, Saisai and Yang, Jianyi and Ramsundar, Bharath and Pande, Vijay S},
  journal={{ACS Cent. Sci.}},
  volume={4},
  number={11},
  pages={1520--1530},
  year={2018},
  publisher={ACS Publications}
}

@article{lim2019predicting,
  title={{Predicting drug--target interaction using a novel graph neural network with 3D structure-embedded graph representation}},
  author={Lim, Jaechang and Ryu, Seongok and Park, Kyubyong and Choe, Yo Joong and Ham, Jiyeon and Kim, Woo Youn},
  journal={J. Chem. Inf. Model.},
  volume={59},
  number={9},
  pages={3981--3988},
  year={2019},
  publisher={ACS Publications}
}

@article{moon2022pignet,
  title={{PIGNet: a physics-informed deep learning model toward generalized drug--target interaction predictions}},
  author={Moon, Seokhyun and Zhung, Wonho and Yang, Soojung and Lim, Jaechang and Kim, Woo Youn},
  journal={Chem. Sci.},
  volume={13},
  number={13},
  pages={3661--3673},
  year={2022},
  publisher={Royal Society of Chemistry}
}

@article{gaieb2018d3r,
  title={{D3R Grand Challenge 2: blind prediction of protein--ligand poses, affinity rankings, and relative binding free energies}},
  author={Gaieb, Zied and Liu, Shuai and Gathiaka, Symon and Chiu, Michael and Yang, Huanwang and Shao, Chenghua and Feher, Victoria A and Walters, W Patrick and Kuhn, Bernd and Rudolph, Markus G and others},
  journal={J. Comput. Aided Mol. Des.},
  volume={32},
  number={1},
  pages={1--20},
  year={2018},
  publisher={Springer}
}

@article{parks2020d3r,
  title={D3R grand challenge 4: blind prediction of protein--ligand poses, affinity rankings, and relative binding free energies},
  author={Parks, Conor D and Gaieb, Zied and Chiu, Michael and Yang, Huanwang and Shao, Chenghua and Walters, W Patrick and Jansen, Johanna M and McGaughey, Georgia and Lewis, Richard A and Bembenek, Scott D and others},
  journal={J. Comput. Aided Mol. Des.},
  volume={34},
  number={2},
  pages={99--119},
  year={2020},
  publisher={Springer}
}

@book{gubernator1998structure,
  title={Structure-based ligand design},
  author={Gubernator, Klaus and B{\"o}hm, Hans-Joachim and Mannhold, Raimund and Kubinyi, Hugo and Timmerman, Hendrik},
  year={1998},
  publisher={Wiley Online Library}
}

@article{jiang2021interactiongraphnet,
  title={{InteractionGraphNet: A novel and efficient deep graph representation learning framework for accurate protein--ligand interaction predictions}},
  author={Jiang, Dejun and Hsieh, Chang-Yu and Wu, Zhenxing and Kang, Yu and Wang, Jike and Wang, Ercheng and Liao, Ben and Shen, Chao and Xu, Lei and Wu, Jian and others},
  journal={J. Med. Chem.},
  volume={64},
  number={24},
  pages={18209--18232},
  year={2021},
  publisher={ACS Publications}
}

@article{anderson2003process,
  title={The process of structure-based drug design},
  author={Anderson, Amy C},
  journal={{Chem. Biol.}},
  volume={10},
  number={9},
  pages={787--797},
  year={2003},
  publisher={Elsevier}
}

@article{ding2013discovery,
  title={{Discovery of RG7388, a potent and selective p53--MDM2 inhibitor in clinical development}},
  author={Ding, Qingjie and Zhang, Zhuming and Liu, Jin-Jun and Jiang, Nan and Zhang, Jing and Ross, Tina M and Chu, Xin-Jie and Bartkovitz, David and Podlaski, Frank and Janson, Cheryl and others},
  journal={J. Med. Chem.},
  volume={56},
  number={14},
  pages={5979--5983},
  year={2013},
  publisher={ACS Publications}
}

@article{zhang2022efficient,
  title={{Efficient and Accurate Physics-aware Multiplex Graph Neural Networks for 3D Small Molecules and Macromolecule Complexes}},
  author={Zhang, Shuo and Liu, Yang and Xie, Lei},
  journal={arXiv preprint arXiv:2206.02789},
  year={2022}
}

@article{moesser2022protein,
  title={{Protein-Ligand Interaction Graphs: Learning from Ligand-Shaped 3D Interaction Graphs to Improve Binding Affinity Prediction}},
  author={Moesser, Marc A and Klein, Dominik and Boyles, Fergus and Deane, Charlotte M and Baxter, Andrew and Morris, Garrett M},
  journal={bioRxiv preprint bioRxiv:2022.03.04.483012}, 
  year={2022},
  publisher={Cold Spring Harbor Laboratory}
}

@article{hermosilla2020intrinsic,
  title={{Intrinsic-extrinsic convolution and pooling for learning on 3D protein structures}},
  author={Hermosilla, Pedro and Sch{\"a}fer, Marco and Lang, Mat{\v{e}}j and Fackelmann, Gloria and V{\'a}zquez, Pere Pau and Kozl{\'\i}kov{\'a}, Barbora and Krone, Michael and Ritschel, Tobias and Ropinski, Timo},
  journal={arXiv preprint arXiv:2007.06252},
  year={2020}
}

@article{eismann2021hierarchical,
  title={{Hierarchical, rotation-equivariant neural networks to select structural models of protein complexes}},
  author={Eismann, Stephan and Townshend, Raphael JL and Thomas, Nathaniel and Jagota, Milind and Jing, Bowen and Dror, Ron O},
  journal={Proteins: Struct., Funct., Bioinf.},
  volume={89},
  number={5},
  pages={493--501},
  year={2021},
  publisher={Wiley Online Library}
}

@article{jimenez2017deepsite,
  title={{DeepSite: protein-binding site predictor using 3D-convolutional neural networks}},
  author={Jim{\'e}nez, Jos{\'e} and Doerr, Stefan and Mart{\'i}nez-Rosell, Gerard and Rose, Alexander S and De Fabritiis, Gianni},
  journal={Bioinformatics},
  volume={33},
  number={19},
  pages={3036--3042},
  year={2017},
  publisher={Oxford University Press}
}

@article{moller2022translating,
  title={{Translating from proteins to ribonucleic acids for ligand-binding site detection}},
  author={M{\"o}ller, Lukas and Guerci, Lorenzo and Isert, Clemens and Atz, Kenneth and Schneider, Gisbert},
  journal={Mol. Inform.},
  publisher={Wiley Online Library},
  pages = {2200059},
  volume = {41},
  year={2022},
}

@article{sverrisson2021fast,
  title={{Fast end-to-end learning on protein surfaces}},
  author={Sverrisson, Freyr and Feydy, Jean and Correia, Bruno E and Bronstein, Michael M},
  journal={Proc. IEEE Comput. Soc. Conf. Comput. Vis. Pattern Recognit.},
  pages={15272--15281},
  year={2021}
}

@article{gainza2020deciphering,
  title={{Deciphering interaction fingerprints from protein molecular surfaces using geometric deep learning}},
  author={Gainza, Pablo and Sverrisson, Freyr and Monti, Frederico and Rodola, Emanuele and Boscaini, D and Bronstein, MM and Correia, BE},
  journal={Nat. Methods},
  volume={17},
  number={2},
  pages={184--192},
  year={2020},
  publisher={Nature Publishing Group},
  addendum={\newline \textbf{* The MaSIF approach uses geodesic convolutions on the protein surface to translate chemical and geometric surface features into a numerical vector for downstream applications, such as pocket classification or binding site prediction.}},
}

@article{dai2021protein,
  title={{Protein interaction interface region prediction by geometric deep learning}},
  author={Dai, Bowen and Bailey-Kellogg, Chris},
  journal={Bioinformatics},
  volume={37},
  number={17},
  pages={2580--2588},
  year={2021},
  publisher={Oxford University Press}
}

@article{fout2017protein,
  title={{Protein interface prediction using graph convolutional networks}},
  author={Fout, Alex and Byrd, Jonathon and Shariat, Basir and Ben-Hur, Asa},
  journal={Advances in Neural Information Processing Systems (NeurIPS)},
  volume={30},
  year={2017}
}

@article{morehead2021geometric,
  title={{Geometric Transformers for Protein Interface Contact Prediction}},
  author={Morehead, Alex and Chen, Chen and Cheng, Jianlin},
  journal={arXiv preprint arXiv:2110.02423},
  year={2021}
}

@article{tubiana2022scannet,
  title={{ScanNet: An interpretable geometric deep learning model for structure-based protein binding site prediction}},
  author={Tubiana, J{\'e}r{\^o}me and Schneidman-Duhovny, Dina and Wolfson, Haim J},
  journal={Nat. Methods},
  pages={1--10},
  year={2022},
  volume={19},
  publisher={Nature Publishing Group},
}

@article{sverrisson2022physics,
  title={{Physics-informed deep neural network for rigid-body protein docking}},
  author={Sverrisson, Freyr and Feydy, Jean and Southern, Joshua and Bronstein, Michael M and Correia, Bruno},
  journal={International Conference on Learning Representations (ICLR) Machine Learning for Drug Discovery},
  volume={10},
  year={2022}, 
  pages={1-13}
}

@article{mendez2021geometric,
  title={{A geometric deep learning approach to predict binding conformations of bioactive molecules}},
  author={M{\'e}ndez-Lucio, Oscar and Ahmad, Mazen and del Rio-Chanona, Ehecatl Antonio and Wegner, J{\"o}rg Kurt},
  journal={Nat. Mach. Intell.},
  volume={3},
  number={12},
  pages={1033--1039},
  year={2021},
  publisher={Nature Publishing Group},
}

@article{volkov2022frustration,
  title={{On the Frustration to Predict Binding Affinities from Protein--Ligand Structures with Deep Neural Networks}},
  author={Volkov, Mikhail and Turk, Joseph-Andr{\'e} and Drizard, Nicolas and Martin, Nicolas and Hoffmann, Brice and Gaston-Math{\'e}, Yann and Rognan, Didier},
  journal={J. Med. Chem.},
  volume={65}, 
  pages={7946-7958},
  year={2022},
  publisher={ACS Publications}
}

@article{wang2004pdbbind,
  title={{The PDBbind database: Collection of binding affinities for protein- ligand complexes with known three-dimensional structures}},
  author={Wang, Renxiao and Fang, Xueliang and Lu, Yipin and Wang, Shaomeng},
  journal={J. Med. Chem.},
  volume={47},
  number={12},
  pages={2977--2980},
  year={2004},
  publisher={ACS Publications}
}

@article{cheng2020design,
  title={{The design and development of covalent protein-protein interaction inhibitors for cancer treatment}},
  author={Cheng, Sha-Sha and Yang, Guan-Jun and Wang, Wanhe and Leung, Chung-Hang and Ma, Dik-Lung},
  journal={J. Hematol. Oncol.},
  volume={13},
  number={1},
  pages={1--14},
  year={2020},
  publisher={BioMed Central}
}

@article{schreiber2021rise,
  title={{The rise of molecular glues}},
  author={Schreiber, Stuart L},
  journal={Cell},
  volume={184},
  number={1},
  pages={3--9},
  year={2021},
  publisher={Elsevier}
}

@article{li2022protacs,
  title={{PROTACs: past, present and future}},
  author={Li, Ke and Crews, Craig M},
  journal={Chem. Soc. Rev.},
  year={2022},
  volume={51}, 
  pages={5214-5236},
  publisher={Royal Society of Chemistry}
}

@article{salton2016small,
  title={{Small molecule modulators of pre-mRNA splicing in cancer therapy}},
  author={Salton, Maayan and Misteli, Tom},
  journal={Trends Mol. Med.},
  volume={22},
  number={1},
  pages={28--37},
  year={2016},
  publisher={Elsevier}
}

@article{hahn2021best,
  title={{Best practices for constructing, preparing, and evaluating protein-ligand binding affinity benchmarks}},
  author={Hahn, David F and Bayly, Christopher I and Macdonald, Hannah E Bruce and Chodera, John D and Mey, Antonia SJS and Mobley, David L and Benito, Laura Perez and Schindler, Christina EM and Tresadern, Gary and Warren, Gregory L},
  journal={arXiv preprint arXiv:2105.06222},
  year={2021}
}

@article{hoogeboom2022equivariant,
  title={{Equivariant diffusion for molecule generation in 3D}},
  author={Hoogeboom, Emiel and Satorras, V{\i}́ctor Garcia and Vignac, Cl{\'e}ment and Welling, Max},
  journal={International Conference on Machine Learning (ICML)},
  volume={39},
  pages={8867--8887},
  year={2022},
  organization={PMLR}
}

@article{corso2022diffdock,
  title={{DiffDock: Diffusion Steps, Twists, and Turns for Molecular Docking}},
  author={Corso, Gabriele and St{\"a}rk, Hannes and Jing, Bowen and Barzilay, Regina and Jaakkola, Tommi},
  journal={arXiv preprint arXiv:2210.01776},
  year={2022}, 
addendum={\newline \textbf{** DiffDock frames molecular docking as a generative task and uses diffusion-based generative models to obtain ligand poses and confidence estimates. It shows substantially improved performance over existing state-of-the-art docking methods on a commonly used benchmark while achieving relatively quick run times.}}
}

@article{anonymous2023structurebased,
    title={{Structure-based Drug Design with Equivariant Diffusion Models}},
    author={Anonymous},
    year={2023},
    pages={1-13},
    journal={https://openreview.net/forum?id=uKmuzIuVl8z},
    addendum={\newline \textbf{* DiffSBDD introduces a generative diffusion model that designs molecules directly in the 3D environment of the active binding site of a protein by denoising normally distributed sets of points.}}, 
    note={under review}
}

@article{igashov2022equivariant,
  title={Equivariant 3D-Conditional Diffusion Models for Molecular Linker Design},
  author={Igashov, Ilia and St{\"a}rk, Hannes and Vignac, Cl{\'e}ment and Satorras, Victor Garcia and Frossard, Pascal and Welling, Max and Bronstein, Michael and Correia, Bruno},
  journal={arXiv preprint arXiv:2210.05274},
  year={2022}
}

@article{
    anonymous2023d,
    title={{3D Equivariant Diffusion for Target-Aware Molecule Generation and Affinity Prediction}},
    author={Anonymous},
    year={2023},
    pages={1-13},
    journal={https://openreview.net/forum?id= kJqXEPXMsE0},
    note={under review}
}
\end{multicols}
\end{spacing}
\end{document}